\documentclass[prl,twocolumn,amsmath,amssymb]{revtex4}

\usepackage{graphicx,epsfig,color}

\newcommand{\bfr}{ {\boldsymbol r} }
\newcommand{\dint}{ {\mathrm d} }
\newcommand{\half}{\frac{1}{2}}
\newcommand{\HH}{\mathcal{H}}

\newcommand{\DD}{\mathcal{D}}
\newcommand{\UU}{\mathcal{U}}
\newcommand{\OO}{\mathcal{O}}
\newcommand{\del}{\partial}
\newcommand{\lbra}{\left\langle}
\newcommand{\rket}{\right\rangle}

\newcommand{\kB}{k_{\rm B}}
\newcommand{\kernel}{\hat{K}}

\newcommand{\sD}{{\rm D}}
\newcommand{\sQ}{{\rm Q}}

\newcommand{\bC}{{\bar{C}}}

\begin{document}

\title{Effective field theory approach to  Casimir interactions on soft matter surfaces}

\author{Cem Yolcu, Ira Z. Rothstein and Markus Deserno} 
\affiliation{%
  Department of Physics, Carnegie Mellon University, 5000 Forbes Ave.,
  Pittsburgh, PA 15213, USA }%

\date{\today}

\begin{abstract}
  We utilize an effective field theory approach to calculate Casimir
  interactions between objects bound to thermally fluctuating fluid
  surfaces or interfaces.  This approach circumvents the complicated
  constraints imposed by such objects on the functional integration
  measure by reverting to a point particle representation.  To capture
  the finite size effects, we perturb the Hamiltonian by 
  $\Delta \HH$ that encapsulate the particles' response to external
  fields.  $\Delta \HH$ is systematically expanded in a series of
  terms, each of which scales homogeneously in the two power counting
  parameters: $\lambda\equiv R/r$ , the ratio of the typical object
  size ($R$) to the typical distance between them ($r$), and $\delta
  \equiv \kB T/k$, where $k$ is the modulus characterizing the surface
  energy.  The coefficients of the terms in $\Delta \HH$ correspond to
  generalized polarizabilities and thus the formalism applies to rigid
  as well as deformable objects.  Singularities induced by the point
  particle description can be dealt with using standard renormalization
  techniques.  We first illustrate and verify our
  approach by re-deriving known pair forces between
  circular objects bound to films or membranes.
  To demonstrate its efficiency and versatility, we then
  derive a number of new results:  The triplet interactions present
  in these systems, a higher order correction to the film interaction,
  and general scaling laws for the leading order interaction valid for
  objects of arbitrary shape and internal flexibility.
\end{abstract}

\maketitle 

Objects which constrain a fluctuating field experience a Casimir interaction
\cite{Casimir}. The underlying fluctuations can be either quantum
mechanical or thermal in origin \cite{RevModPhys}.
In this letter we will be interested in forces induced by thermal
fluctuations between particles bound to surfaces characterized by
surface tension (films) \cite{LO, NO, LiKardar, GGK} or bending rigidity
(membranes) \cite{GPB, GGK, PLub, Fournier}.

The non-trivial aspect of such calculations tends to arise from the
constraints that the extended objects impose on the
partition sum.  This issue is usually dealt with by pinning the field
to the surface of the objects through delta-functions in the
integration measure \cite{LiKardar}. A clear exposition of this
method, applied to compact objects in fluid membranes and films, can
be found in Ref. \cite{GGK}. For the electromagnetic Casimir effect
this approach was recently improved and systematized within
the framework of scattering theory, where the
constraints of the objects enter the interaction energy through their
scattering matrix coefficients \cite{EmigPRL}.

In this paper we employ an effective field theory (EFT) formalism,
orignally developed to study the gravity wave profile for inspiralling
black holes \cite{NRGR}, to streamline the boundary condition issue.
This formalism has been utilized to derive not only new results in
gravitational wave physics \cite{PRR} but also to calculate the
leading order finite size correction to the Abraham-Dirac-Lorentz
radiation reaction force law in classical electrodynamics \cite{GLR}.
Both of the these applications dealt with classical
non-fluctuating fields, whereas here we will generalize the
formalism to allow for finite temperatures.

To illustrate the effective field theory approach to soft matter
surfaces most transparently, we will focus on constraints imposed by
mobile but rigid objects which pin field fluctuation modes along
their circumference. Extensions towards more general types of
constraints entails no change in formalism.

We will assume that the bare surface Hamiltonian $\HH$ is a quadratic
functional of some field $\phi(\bfr)$:
\begin{equation}
  \HH[\phi] = \half \, k \! \int\dint^2r\,\phi\kernel\phi
  \ , \label{eq:H}
\end{equation}
where $k$ is a generalized modulus and the kernel $\kernel$ defines
the physics of the problem. For films $\kernel=-\nabla^2$ while for
membranes $\kernel=(\nabla^2)^2$. In both cases $\phi(\bfr)$ is the
surface height in Monge parametrization \cite{Safran}.

The idea behind the EFT formalism is to treat the objects as point
particles and to recapture their internal structure through additional terms
$\Delta \HH = \sum_a C_a O_a$ in the Hamiltonian, 
where the scalars $O_a$ are polynomial in the field and its
derivatives and $a$ labels the particles.
The coefficients $C_a$ are chosen to reproduce the long
wavelength physics, in analogy with block spin renormalization, and
are fixed by a matching calculation.  In principle one must add all
terms which are consistent with the underlying theory.  On dimensional
grounds the coefficients of these new terms will scale with powers of
the particle size, $R$, such that the limit $R\rightarrow 0$ is well
defined.  Let us for the moment assume that the new terms obey a shift
symmetry $\Delta \HH(\phi)= \Delta \HH(\phi + h)$, where $h$ is a constant.
This symmetry eliminates boundary conditions in which the object is pinned
to a fixed height.  Violating this symmetry leads to long wavelength
fluctuations which can lead to pathologies.  Given this restriction,
it is clear that we may truncate the sum in a derivative expansion, as
each derivative will scale as $\lambda\equiv R/r$.  Quadratic terms in
$\phi$ generate multipoles when the particles are subjected to
external fields, hence we may interpret their coefficients as
polarizabilities. Terms which are higher order in the fields will not
have such a simple interpretation.  Such terms are suppressed by
powers of $\delta \equiv \kB T/k$. To see this, we may simply rescale
the field $\phi\rightarrow\sqrt{\delta}\,\phi$. In this way the leading term
in the partition function has a well defined $\delta \rightarrow 0$ limit
and non-linearities are automatically suppressed by powers of
$\delta$. These $\delta$-corrections will be treated in a
forthcoming publication. Here we will be only concerned with the more
canonical $\lambda$-corrections.

The leading order in $\lambda$ is unique and $\OO(\lambda^2)$:
\begin{equation}
  \Delta \HH^{(2)} =\half\sum_a C^a_{ij} \left[ \del_{i}\phi(\bfr_a)
    \del_{j}\phi(\bfr_a) \right] \ . \label{eq:Heff}
\end{equation}
The tensorial polarizability $C^a_{ij}$ allows for non-axi\-sym\-met\-ric
objects, but for the sake of simplicity we will restrict to the symmetric case $C^a_{ij}=C^\sD_a \delta_{ij}$.
The coefficient $C^\sD_a$, which can be fixed by treating a single object in isolation, is the
isotropic \emph{dipole polarizability}.  The reason for this
nomenclature will become clear once we match for $C^\sD_a$.  Note that
in the EFT formalism determining the proper Hamiltonian to reproduce
the long distance physics is both conceptually and technically
independent of finding the associated force between objects. We will
therefore first discuss how to compute Casimir interactions between
polarizable objects, as described by the Hamiltonian $\HH_{\rm eff}=\HH+\Delta\HH$,
and afterwards explain how the polarizabilities are fixed by
a matching procedure.

The canonical partition function of the system described by the
effective Hamiltonian $\HH_{\rm eff}$ is given by 
\begin{equation}
  Z=\int\DD\phi \; {\rm e}^{-\beta\,\HH_{\rm eff}} = Z_0 \lbra
  {\rm e}^{-\beta\Delta\HH} \rket \ , \label{eq:PF}
\end{equation}
where $Z_0$ is the partition function of the free Hamiltonian
($\Delta\HH=0$) and $\langle\cdots\rangle$ denotes the associated
Gaussian average. Substituting $\Delta\HH$ from its definition in
Eqn.~(\ref{eq:Heff}), one obtains the free energy of interaction,
$\UU$, through
\begin{eqnarray}
  -\beta\,\UU & = & \log(Z/Z_0) \; = \; \log\left\langle 
    e^{-\beta\Delta\HH}\right\rangle \nonumber \\
  & = &  \sum_{n=1}^{\infty} \frac{1} {n!}  \lbra \left( -
    \frac{\beta}{2} \sum_a C^\sD_a \left[ \del_i \phi(\bfr_a) \right]^2 
  \right)^n\rket_{\rm c} \ . \;\;\;\; \label{eq:freeEnergy}
\end{eqnarray}
This cumulant expansion can be represented as a series of diagrams
\cite{Binney}, each depicting a (connected) $2n$-point function of
some derivatives of the field.

The expansion~(\ref{eq:freeEnergy}) encodes $n$-body contributions to
the free energy from the $n^{\rm th}$ term onwards.  For a given
$n$-body force there are subleading corrections stemming from higher
multipoles as well as terms non-linear in the lower multipole polarizabilities.

\begin{figure}[thb]
\centering
\includegraphics[scale=.29]{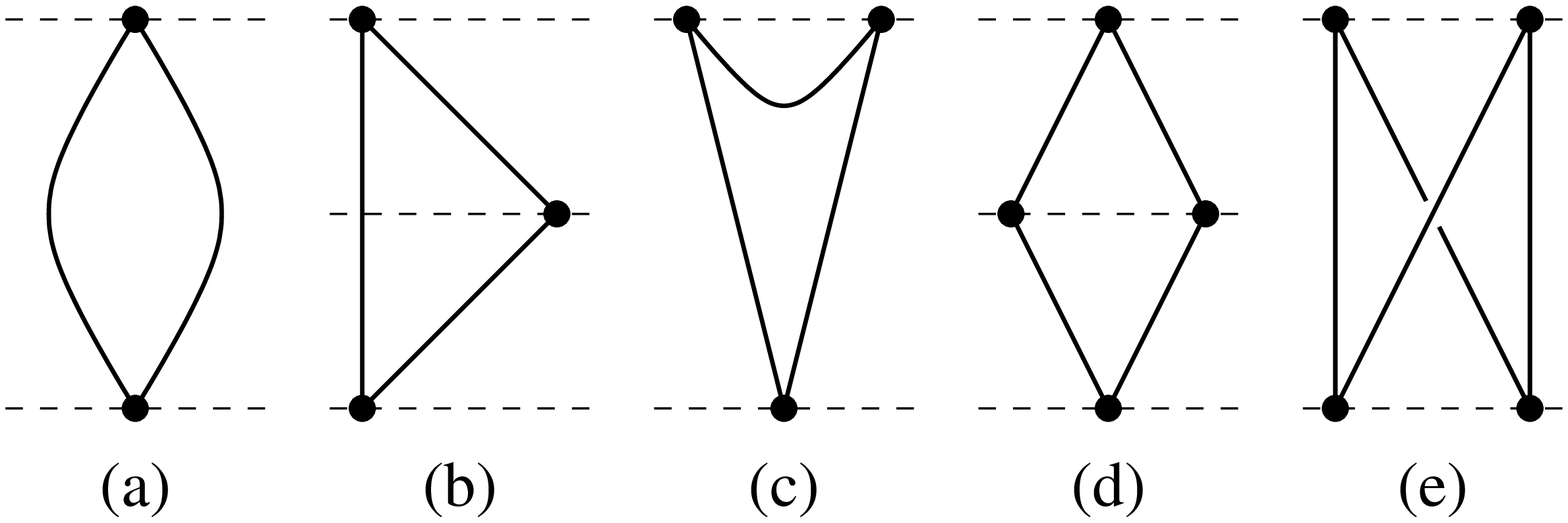}
\caption{Relevant Feynman diagrams for the calculations in this
  letter. Solid lines correspond to propagators, as usual, while dashed
  lines represent world lines of particles.}
\label{fig:diagrams}
\end{figure}

Let us now illustrate this formalism by applying it to some cases of
interest.  We first consider circular particles on a surface with
tension-dominated energy density, \emph{i.e.} a film. The relevant
differential operator is then $-\nabla^2$, with Green function
$G(\bfr,\bfr')=-\frac{1}{2\pi}\log \lvert \bfr-\bfr' \rvert$, and the
modulus $k$ will be replaced by the more familiar $\sigma$ for surface
tension. At $\OO(\lambda^4)$ we have the diagram in Fig.~\ref{fig:diagrams}(a),
which represents the lowest order pair interaction. It Wick-contracts to
\begin{equation}
  -\beta\,\UU^{(4)} = \sum_{a,b} \frac{C^\sD_a C^\sD_b}{4\sigma^2}
  \left[ \del_i\del_j G(\bfr_a,\bfr_b) \right]^2
    \label{eq:dipdip_film_unmatched}
\end{equation}
and evaluates to the pair potential
\begin{equation}
-\beta\,\UU^{(4)}_{\{ab\}} = \frac{C^\sD_aC^\sD_b}{4\pi^2 \sigma^2
  r_{ab}^4} \ , \label{dipdip}
\end{equation}
where we define $\bfr_{ab}=\bfr_a-\bfr_b$ and denote by $\{\ldots\}$
an $n$-tuplet of distinct particles.  Self energies corresponding to
self-links in diagrams, such as the one in
Fig.~\ref{fig:diagrams}(c), lead to divergent contributions to the
interaction energy. However, for derivative interactions these
divergences are all power like and can be absorbed into
$C^\sD_a$. These divergences carry no physical information, as there
is no non-trivial renormalization group flow, and thus effectively we
may set these diagrams to zero.

Let us now consider contributions beyond $\OO(\lambda^4)$ to the two
body interaction, for which there are two possible sources: Higher
order  corrections to $\Delta \HH$
itself and higher order terms in the cumulant
expansion.  The next order term in $\Delta \HH$, which is not
subleading in $\delta$, is given by
\begin{equation}
\label{quad}
\Delta \HH^{(4)} =\half\sum_a C^\sQ_a
  \left[ \del_{i}\del_{j}\phi(\bfr_a) \right]^2 .
\end{equation}
It corresponds to a quadrupole polarizability and is $\OO(\lambda^4)$.
In principle there is also a term involving
$\del_i\del_i\phi(\bfr_a)$, but it can be removed by a redefinition of
$\phi$ \cite{TASI}, owing to the fact that
  $\del_i\del_i\phi = 0$ is the Euler-Lagrange equation for the
  problem. Inserting (\ref{eq:Heff})$+$(\ref{quad}) into the cumulant
expansion (\ref{eq:freeEnergy}) generates a quadrupole-dipole
interaction at $\OO(\lambda^6)$, given by
\begin{equation}
-\beta \, \UU_{\{ab\}}^{(6)} =
\frac{2(C^\sQ_aC^\sD_b+C^\sD_aC^\sQ_b)}{\pi^2\sigma^2 r_{ab}^6}
\ . \label{quaddip}
\end{equation}
At $\OO(\lambda^6)$ one might additionally expect a contribution
involving three dipole interactions in the cumulant expansion.
However these vanish, since such terms necessarily
  involve self-energies, {\em c.f.} diagram~\ref{fig:diagrams}(c).

At $\OO(\lambda^8)$ we expect the quadrupole-quadrupole interaction,
arising from $[\del^4 G(\bfr_a,\bfr_b)]^2$,
but also a non-linear dipole-dipole term, proportional
to four dipole polarizabilities and stemming from the $4^{\rm th}$ cumulant,
and diagram \ref{fig:diagrams}(e), with $[\del^2 G(\bfr_a,\bfr_b)]^4$:
\begin{equation}
-\beta \, \UU_{\{ab\}}^{(8)} =
\left[\frac{36\,C^\sQ_aC^\sQ_b}{\pi^2\sigma^2} +
  \frac{(C^\sD_a)^2(C^\sD_b)^2}{32 \pi^4\sigma^4}\right]
\frac{1}{r_{ab}^8} \ .
 \label{order-8}
\end{equation}
As long as the objects resist curvature, $C^\sQ_a \ne 0$.
However, if in addition to vertical  translations the objects
can also tilt, their ability to align with local
gradients in $\phi$ implies $C^\sD_a=0$ and the lowest order interaction is $\OO(\lambda^8)$.
While the precise value of $C^\sQ_a$ might be hard to calculate for arbitrarily shaped rigid or elastic
objects, Eqns.~(\ref{dipdip},\ref{quaddip},\ref{order-8}) nevertheless
show that on films these will always interact with an asymptotic
$r^{-8}$ potential, thus generalizing a finding obtained in Ref.~\cite{NO} for ellipsoidal objects.
If the objects are identical, $(C^\sQ)^2>0$ implies they attract.

The lowest  order $3$-body interaction arises from three
dipole insertions.  Henceforth using the shorthand notation
$G^{ab}=G(\bfr_a,\bfr_b)$ and denoting partial derivatives by
subscripts, the relevant interaction is
\begin{equation}
-\beta\,\UU_{\{abc\}}^{(6)} = -\frac{C^\sD_aC^\sD_bC^\sD_c}{\sigma^3}
G_{ij}^{ab} G_{jk}^{bc}G_{ki}^{ca} \ ,  \label{eq:3bodyFilm1}
\end{equation}
which corresponds to diagram~\ref{fig:diagrams}(b) and yields a
triplet interaction that scales as $\left(
r_{ab}r_{bc}r_{ca}\right)^{-2}$.  However, owing to the symmetries of
the tensor $G_{ij}=(\delta_{ij}-2\hat{r}_i\hat{r}_j)/(2\pi r^2)$,
${\rm Tr}[G^n]=0$ if $n$ is odd.  Thus the leading dipole contribution
to any $n$-body interaction vanishes for $n$ odd.  The first
non-vanishing 3-body interaction therefore arises at
$\OO(\lambda^8)$, 
from terms which are quadratic in one of the dipole polarizabilities,
diagram \ref{fig:diagrams}(d):
\begin{equation}
 - \beta \, \UU_{\{abc\}}^{(8)} =
 \frac{C^\sD_aC^\sD_bC^\sD_c}{16\pi^4\sigma^4} \left( \frac{C^\sD_a}
      {r_{ab}^4r_{ac}^4} + \frac{C^\sD_b} {r_{ba}^4r_{bc}^4} +
      \frac{C^\sD_c} {r_{cb}^4r_{ca}^4} \right).
  \label{eq:3bodyFilm2}
\end{equation}
This term drives triplets to \emph{attract}, irrespective of their
relative placement, as long as $C^\sD>0$.  A possible
dipole-quadrupole-dipole interaction also scales as $\OO(\lambda^8)$,
but again the symmetries of $G_{ij}$ and $G_{ijkl}$ force it to
vanish.

So far we have calculated the forces in terms of a set of
polarizability coefficients.  These are fixed by calculating an
observable in the full (finite-sized particle) theory, expanding it in
powers of $\lambda$, and choosing the polarizabilities such that the
EFT reproduces the result to the appropriate order in $\lambda$.  The
point to emphasize is that we are free to fix these coefficients by
any means we choose.  Thus it behooves us to choose as simple a
setting as possible: we will place a single
particle in a simple stationary external field where we can
easily calculate its response.

To illustrate the procedure, let us match for $C^\sD_a$ in the case of
a rigid horizontal inclusion on a film.  Consider placing the point
particle in a background field such that the total field is given by
$\delta\phi(\bfr)+\phi_{\rm bg}(\bfr)$, where $\delta \phi(\bfr)$ is
the induced field generated by the polarization of the inclusion.
Terms linear in $\delta\phi $ in the Hamiltonian correspond to induced
point sources
\begin{equation}
  \rho_a(\bfr) = -C^\sD_a \del_i \left[ \delta(\bfr-\bfr_a) \del_i
    \phi_{\rm bg}(\bfr)\right] \label{source}
\end{equation}
in the effective theory. The field emitted by an induced source is
thus given by
\begin{equation}
  \phi_a (\bfr) = -\frac{C^\sD_a}{k} \del_{i}^{(a)}
  G(\bfr,\bfr_a) \del_{i}^{(a)} \phi_{\rm bg}
  (\bfr_a) \ . \label{eq:response}
\end{equation}
We may pick any background field we wish, but clearly it is
simplest to choose the lowest multipole field configuration necessary
to generate a non-zero response. For a rigid horizontal inclusion this
is a dipole field $\phi^{\rm bg} \sim r \cos\varphi$,
\emph{i.e.} one of constant slope.  After solving the elementary
boundary value problem (BVP) in the full theory and comparing to the
effective theory result, one finds $C^\sD=2\pi R^2\sigma$.  To match
for $C^\sQ$, we need a background with curvature, so we choose
$\phi^{\rm bg} \sim r^2 \cos(2 \varphi)$. Repeating the exercise we just
performed, after appropriately generalizing (\ref{source}) to account
for the two derivative nature of (\ref{quad}), gives $C^\sQ=
\frac{\pi}{2} R^4 \sigma$.  By
Eqns.~(\ref{dipdip},\ref{quaddip},\ref{order-8}) this yields the
 pair interaction up to $\OO(\lambda^8)$:
\begin{equation}
 -\beta \, \UU_{\{ab\}}^{(\le 8)} = \frac{R^4}{r_{ab}^4} +
 4\frac{R^6}{r_{ab}^6} +
 \left(9+\frac{1}{2}\right)\frac{R^8}{r_{ab}^8} \ .
  \label{eq:pairFilmNumbers}
\end{equation}
The $r_{ab}^{-4}$ (dipole-dipole) and $r_{ab}^{-6}$
(dipole-quadrupole) interactions shown above agree with those derived in
Ref.~\cite{LO}. However, the $r_{ab}^{-8}$ term consists of a lowest
order quadrupole-quadrupole piece (prefactor ``$9$'', also given in
Ref.~\cite{LO}) plus a non-linear dipole-dipole correction (prefactor
``$1/2$'') that has not been previously calculated. Recall that if the discs can also tilt,
every term proportional to a dipole polarizability will vanish and thus only
the interaction $-9\,\kB T (R/r_{ab})^8$ will survive.
As discussed before, a $r_{ab}^{-8}$-term will remain even for non-circular or
bendable objects; only its prefactor will be different, owing to the
associated BVP being slightly different.

Observe that the surface tension $\sigma$ (or generally the modulus
$k$) cancels from the final result, because in every
  term of $-\beta\,\UU$ the number of Green functions,
each accompanied by a factor $\sigma^{-1}$, always
matches the number of polarizabilities, each
  proportional to $\sigma$.  This will no longer be the case once
corrections beyond lowest order in $\delta$ are included.

Now consider the case of particles embedded in a surface with a
bending-dominated energy density, \emph{i.e.} a membrane. The kernel
of the Hamiltonian in this case is $\kernel=(\nabla^2)^2$, with the
Green function $G(\bfr,\bfr') = \frac{1}{8\pi} \lvert \bfr-\bfr'
\rvert^2 \log \lvert \bfr-\bfr' \rvert$, and the generic modulus $k$
will be replaced by the bending rigidity $\kappa$.  We will assume
that the particles can adjust to a constant slope background by
tilting, so the first non-vanishing polarizability will be
quadrupole in nature. Observe that this time, at $O(\lambda^2)$, we
need to write down two distinct terms:
\begin{equation}
\Delta\HH = \half \sum_a \left[ \bC^\sQ_a (\del_i \del_j
  \phi)^2(\bfr_a) + \bC^{\sQ'}_a (\del_i\del_i\phi)^2(\bfr_a) \right] \ .
\end{equation}
Terms involving the Laplacian cannot be removed via a field
re-definition, because $\del_i\del_i\phi=0$ is no longer the
Euler-Lagrange equation.  Notice that due to the different kernel the
quadrupole polarizabilities $\bC^\sQ$ and $\bC^{\sQ'}$ scale $\sim
R^2$, and not $\sim R^4$ as in the film case.

The pair interaction follows easily from an expression
analogous to Eqn.~(\ref{eq:dipdip_film_unmatched}):
\begin{equation}
-\beta\, \UU_{\{ab\}}^{(4)} = \frac{\bC^\sQ_a
  \bC^\sQ_b}{2\kappa^2} \left(G^{ab}_{ijkl}\right)^2 +
\frac{\bC^{\sQ'}_a\bC^\sQ_b+\bC^\sQ_a\bC^{\sQ'}_b}{2\kappa^2} \left(
G^{ab}_{iikl} \right)^2 \ .
      \label{eq:mem-pair}
\end{equation}
The term $\propto\bC^{\sQ'}_a\bC^{\sQ'}_b (G_{iikk}^{ab})^2$ vanishes
because $G^{ab}_{iikk} = \delta (\bfr_a-\bfr_b)$ by definition of
the biharmonic Green function.  The other contractions are found to be
$(G^{ab}_{ijkl})^2 = 4/(2 \pi r_{ab}^2)^2$ and $(G^{ab}_{iikl})^2 =
2/(2\pi r_{ab}^2)^2$.

The leading 3-body interaction in the membrane case
stems from the third order cumulant.  The
interaction is derived from an expression analogous to
Eqn.~(\ref{eq:3bodyFilm1}) and is given by
\begin{eqnarray}
 - \beta \, \UU_{\{abc\}}^{(6)} & = &
  -\frac{1}{\kappa^3} \bigg[ \bC^\sQ_a \bC^\sQ_b
   \bC^\sQ_c \Big(G^{ab}_{ijkl} G^{bc}_{klmn} G^{ca}_{mnij}\Big)
   \nonumber\\ & & + \sum \bC^{\sQ'} _a \bC^\sQ_b \bC^\sQ_c
   \Big(G^{ab}_{iikl} G^{bc}_{klmn} G^{ca}_{mnjj}\Big)\bigg] \ ,
 \;\;\;\;\;
 \label{eq:3-body_membrane}
\end{eqnarray}
where the sum includes three terms in which the prime accompanies $a$,
$b$, or $c$.  The full expression for this result is rather lengthy,
but it simplifies greatly once we have the relationship between $\bC^\sQ$
and $\bC^{\sQ '}$.

In the case of membranes we may perform a completely analogous
matching procedure to the one described for the film. One finds the
polarizabiliy coefficients $\bC^{\sQ} = 4\pi R^2\kappa$ and
$\bC^{\sQ'} = -\pi R^2 \kappa$, which through Eqn.~(\ref{eq:mem-pair})
then yield the well-known pair potential \cite{GGK, GPB}
\begin{equation}
-\beta \, \UU^{(4)}_{\{ab\}} = 6\;\frac{R^4}{r^4_{ab}} \ .
\end{equation}
A similar force law could be derived for soft objects, after solving the
relevant BVP to extract the polarizabilities. The potential would again
decay $\sim (R/r_{ab})^{4}$, while its strength would depend
on the relative elastic moduli of the objects and the bulk membrane,
approaching $6$ in the limit of rigid objects.

With the polarizabilities fixed, we can simplify the 3-body interaction
(\ref{eq:3-body_membrane}) for the membrane, leaving
\begin{equation}
  -\beta \,\UU_{\{abc\}}^{(6)} = - 4\,(c_{ab} +c_{bc}+c_{ca})
  \frac{R^6}{r_{ab}^2 r_{bc}^2 r_{ca}^2} \ ,
\end{equation} 
where $c_{ab}=\cos[2(\alpha_a-\alpha_b)]$ and $\alpha_a$ denotes the
inner angle of the triangle at point $a$, {\em etc}.  Unlike in the
film-case, the sign of this triplet interaction depends on the
relative orientation of the three particles.  For instance, for an
equilateral triangle of side-length $s$ one finds $\UU^{(6)}_{\rm triplet}(s)=+12\,\kB
T(R/s)^6$, but for an isosceles triangle with angles
$(30^\circ,120^\circ,30^\circ)$ and short side $s$ one finds
$\UU^{(6)}_{\rm triplet}(s)=-\frac{4}{3}\,\kB T(R/s)^6$.

To conclude, in this letter we have utlilized the effective field
theory approach introduced in \cite{NRGR} to streamline calculations
of Casimir forces on fluctuating two dimensional surfaces. We have
reproduced well known results and derived several new ones, pertaining to
non-linear corrections, $3$-body terms, and general scaling laws for
the leading order interaction between objects that are arbitrarily shaped and
possibly flexible. However, this formalism
extends well beyond the basic cases studied here and can be applied
to many further situations in which existing techniques become
rather unwieldy. For instance, the corrections beyond linear order
in $\kB T/k$ follow in a straightforward manner and will be presented
elsewhere. In addition, our results can be utilized to calculate the
forces and torques between non-rigid objects
or phase-segregated surface domains. This does not require
further field-theoretical sophistication but merely the calculation of their
polarizability, which for complicated objects one might even decide
to extract from experiment.

We gratefully acknowledge stimulating discussions with J. Guven and
M. Oettel.  IZR is supported by US DOE contract, 22645.1.1110173.
IZR is thankful to the Caltech high energy theory  group for its hospitality
and to the Gordon and Betty Moore Foundation for support.  MD
would like to thank the Theory Department of the MPI for Polymer
Research for its hospitality.



\end{document}